\documentclass[
a4paper,%
12pt,%
twoside%
]{article}

\usepackage[dvips]{graphicx}

\usepackage{hyperref}

\hypersetup{
unicode=true,
pdfauthor={\textcopyright\ Sergei P. Maydanyuk},
pdfkeywords={
alpha-decay,
multiple internal reflections,
wave packet,
tunneling, times,
penetrability and reflection coefficients}}

\title{Method of multiple internal reflections in description of
$\alpha$-decay of nucleus
\footnote{The paper is published (in Russian) in Journ.
``Problems of atomic science and tehnology''
(RFNC-VNIIEF, Sarov, Russia),
Vol.~1 (2), p.16--19 (2002)}}

\author{
Sergei~P.~Maydanyuk\thanks{E-mail: maidan@kinr.kiev.ua},
Vladislav~S.~Olkhovsky\thanks{E-mail: olkhovsk@kinr.kiev.ua} \,
and
Sergei~V.~Belchikov\thanks{E-mail: belchik@kinr.kiev.ua} \\
\small\emph{
Institute for Nuclear Research,
National Academy of Sciences of Ukraine} \\
\small\emph{prosp. Nauki, 47, Kiev-28, 03680, Ukraine}}

\date{\small{18 Dec 2001}}

\begin{document}
\begin{sloppypar}

\maketitle

\vspace{-10mm}
\begin{abstract}
A spatial-time method of description of $\alpha$-decay of a nucleus
is presented. The method is based on the quantum mechanical model
of the $\alpha$-decay of a compound nuclear system with the
barrier. $\alpha$-decay dynamics is described on the basis of
multiple internal reflections of wave packets from boundaries of
the decay barrier.
The method allows to study this process in details at any time
moment or point of space. According to the method, some time
before the time moment of the nuclear decay the nucleus makes
oscilations. In the calculations of the time duration of the
$\alpha$-decay of the compound nucleus we propose to take into
account the time durations of these oscillations. The method of
the multiple internal reflections allows to calculate the time of
such oscillations before the $\alpha$-decay and the total
$\alpha$-decay time.
\end{abstract}

{\bf PACS numbers:} 03.40.Kf, 03.50.De, 03.65.Nk, 41.20.Jb

{\bf Keywords:}
alpha-decay,
multiple internal reflections,
wave packet,
tunneling, times,
penetrability and reflection coefficients


\vspace{11mm}
An approach for description of a motion of a non-relativistic
particle above a potential barrier in the one-dimensional problem
with taking into account of multiple internal reflections of plane
waves, which describe the propagation of this particle,
between boundaries of this barrier was studied in number of papers
\cite{McVoy.1967.RMPHA,Anderson.1989.AJPIA,Fermor.1966.AJPIA}
and has been known for the long time.
Here, each such wave was considered as stationary, on their basis
only the stationary solutions for wave functions (WF) were obtained
and there was a difficulty for description of tunneling of the
particle under the barrier.

We have generalised this approach, developing the nonstationary
method of solution of the tunneling problem of the non-relativistic
particle through the one-dimensional barrier, which is constructed
with use of the multiple internal reflections of nonstationary
wave packets (WP) between the barrier boundaries
\cite{Maydanyuk.2000.UPJ}\footnote{in \cite{Maydanyuk.2002.JPS}
the extended version of this method is presented, where its proof,
main definitions and peculiarities are given in study of
the simplest problem of tunneling, its application to the
analysis of the particle tunneling through the one-dimensional
barriers of more complicated forms and to the analysis of the
particle in its spherically symmetric scattering upon nucleus
are studied,
tunneling of photons is studied by this method (for the first
time)}.
The non-stationary use of WPs in the processes of the multiple
internal reflections, fulfilled on the basis of papers
\cite{Olkhovsky.1992.PRPLC,Olkhovsky.1997.JPGCE,Maydanyuk.1999.JPS},
allows from the physical point of view to motivate correctely the
approach for tunneling description on the basis of the multiple
internal reflections.
The method allows to study in details the tunneling of the particle
at any point of space and to calculate different times during
tunneling (or propagation).
In calculation of the time durations of tunneling this method has
shown itself as simple and effective enough.

After developing the method formalism for the spherically
symmetrical problems, with its help a problem of elastic scattering
of the non-relarivistic particle upon nucleus was solved and some
new characteristics, describing such process, were obtained
\cite{Maydanyuk.2000.UPJ}.

As a further development of the time analysis of nucleus processes
\cite{Maydanyuk.2000.UPJ,Olkhovsky.1992.PRPLC,Olkhovsky.1997.JPGCE,
Maydanyuk.1999.JPS},
for the first time we present a non-stationary method of the
description of the nyclear $\alpha$-decay on the basis of the
multiple internal reflections of the wave packets relatively on
the barrier boundaries. The method is constructed on the basis of
the quantum mechanical one-particle model of the decay of a
compound nucleus with use of the barrier \cite{Baz.1971} and allows
to fulfill the space-time analysis of the decay process. With its
help, some new stationary and time parameters for the description
of the $\alpha$-decay of the nucleus are found.

Now we describe briefly the method in solution of the one-dimensional
problem of the particle tunneling through the square barrier of the
following form:
\begin{equation}
  V(x) = 
  \left\{
  \begin{array}{crcl}
    0,     & \mbox{ at } & x<0   & \mbox{ (region I)};  \\
    V_{1}, & \mbox{ at } & 0<x<a & \mbox{ (region II)}; \\
    0,     & \mbox{ at } & x>a   & \mbox{ (region III)}.
  \end{array}
  \right.
\label{eq.1}
\end{equation}

Let's assume, that the particle with the mass $m$, falling on the
barrier from the left side and tunneling through it at the energy
level $E$, which locates below than the barrier heights $V_{1}$. A
general stationary solution for WF for this process has the form:
\begin{equation}
  \varphi(x) = 
  \left\{
  \begin{array}{crcl}
    e^{ikx} + A_{R}e^{-ikx},
      & \mbox{ at } & x<0   & \mbox{ (region I)};  \\
    \alpha e^{\xi x} + \beta e^{-\xi x},
      & \mbox{ at } & 0<x<a & \mbox{ (region II)}; \\
    A_{T} e^{ikx},
      & \mbox{ at } & x>a   & \mbox{ (region III)},
  \end{array}
  \right.
\label{eq.2}
\end{equation}
where
$k   = \displaystyle\frac{1}{\hbar} \sqrt{2mE}$,
$\xi = \displaystyle\frac{1}{\hbar} \sqrt{2m(V_{1} - E)}$ and
$\alpha$, $\beta$, $A_{T}$, $A_{R}$ are the normalization factors.

According to the method, we describe tunneling of the particle
under the barrier with use of the WP sequentially by steps of its
transitions through each boundary. In the beginning, on the first
step we consider the particle, which is incident upon the first
boundary of the barrier at point $x=0$ from the left side. This
particle is reflected from the boundary with certain probability
and transits through this boundary with another certain probability,
tunneling inside the barrier further.
One can describe the incident, transitted and reflected
propagations (with taking into account of tunneling) of the
particle relatively of the first boundary of the barrier with use
of the WP of the following form:
\begin{equation}
  \Psi_{i}(x,t) =
    \displaystyle\int\limits_{0}^{+\infty}
    g(E-\bar{E}) \varphi_{i}(k,x) e^{- iEt / \hbar} dE,
\label{eq.3}
\end{equation}
where $\bar{E}$ is the average energy of the particle. The incident,
transmitted and reflected stationary components $\varphi_{i}(x)$
can be found from
\begin{equation}
  \varphi(x) = 
  \left\{
  \begin{array}{crcl}
    e^{ikx} + A_{R}^{0} e^{-ikx}, & \mbox{ at } & x < 0;  \\
    \beta^{0} e^{-\xi x},         & \mbox{ at } & 0 < x < a.
  \end{array}
  \right.
\label{eq.4}
\end{equation}
The constant factors $A_{R}^{0}$ and $\beta^{0}$ are determined
from the continuity conditions of the non-stationary WF
(constructed on the basis of the WP in regions I and II) and its
derivative at point $x = 0$.

On the second step, we consider the particle, tunneling along the
region II and incidenting upon the second boundary of the barrier
at point $x = a$. It transits through this boundarywith with the
certain probability and is reflected from it with the certain
probability also.
One can describe this process with use of the WP of the form
(\ref{eq.3}), where one need to use such stationary WF:
\begin{equation}
  \varphi(x) = 
  \left\{
  \begin{array}{crc}
    \alpha^{0} e^{\xi x} + \beta^{0} e^{-\xi x},
      & \mbox{ at } & 0<x<a; \\
    A_{T}^{0} e^{ikx}, & \mbox{ at } & x>a.
  \end{array}
  \right.
\label{eq.5}
\end{equation}
The constant factors $\alpha^{0}$ and $A_{T}^{0}$ are obtained
from the continuity conditions of the WF and its derivative at
point $x = a$.

On the third step, the particle, reflected from the second boundary
of the barrier, tunnels along the region II back and is incident
upon the first boundary at point $x = 0$. It transits through this
boundary with the certain probability and is reflected from it
with another probability. One can desribe this process with use of
the WP (\ref{eq.3}), where one can use the stationary WF of the
form:
\begin{equation}
  \varphi(x) = 
  \left\{
  \begin{array}{crc}
    \alpha^{0} e^{\xi x} + \beta^{1} e^{-\xi x},
      & \mbox{ at } & 0<x<a; \\
    A_{R}^{1} e^{-ikx}, & \mbox{ at } & x<0.
  \end{array}
  \right.
\label{eq.6}
\end{equation}
The constant factors $\beta^{1}$ and $A_{R}^{1}$ can be calculated
from the continuity conditions of the WF and its derivatives at
point $x = 0$.

Considering further the process of tunneling (and propagating) of
the particle by such a way, we conclude, that any following
step of the particle transition or reflection relatively the
boundary can be reduced to one of two steps, considered above.
For the constant factors one can construct the recurrence
relations.

We describe the propagation and reflection of the particle
relatively the barrier on the basis of total transmitted and
reflected WPs of the form:
\begin{equation}
\begin{array}{lcl}
  \Psi_{tr}(x,t) =
    \displaystyle\int\limits_{0}^{+\infty} g(E-\bar{E})
    \sum A_{T}^{(i)} e^{ikx-iEt / \hbar} dE,
    & \mbox{ at } & x > a;  \\
  \Psi_{ref}(x,t) =
    \displaystyle\int\limits_{0}^{+\infty} g(E-\bar{E})
    \sum A_{R}^{(i)} e^{ikx-iEt / \hbar} dE,
    & \mbox{ at } & x < 0.
\end{array}
\label{eq.7}
\end{equation}
Here, one can calculate the summs of the factors $A_{T}^{(i)}$
and $A_{R}^{(i)}$ on the basis of the recurrence relations found
earlier:
\begin{equation}
\begin{array}{ll}
  \sum\limits_{i=0}^{+\infty} A_{T}^{(i)} =
    \displaystyle\frac{i4k \xi e^{-\xi a-ika}}{F}; &
  F = (k^{2}-\xi^{2})D_{-} + 2ik\xi D_{+}; \\
  \sum\limits_{i=0}^{+\infty} A_{R}^{(i)} =
    \displaystyle\frac{i4k \xi e^{-\xi a-ika}}{F}; &
  D_{\pm} = 1 \pm e^{-2\xi a}.
\end{array}
\label{eq.8}
\end{equation}

All expressions for the constant factors summs, obtained by the
method of multiple internal reflections, coincide with the
corresponding coefficients, calculated with use of standard
stationary approaches. Using the transformation
\begin{equation}
  i\xi \to k_{2},
\label{eq.9}
\end{equation}
where $k_{2} = \displaystyle\frac{1}{\hbar} \sqrt{2m(E-V_{1})}$,
the expressions for the constant factors at each step, the
expressions for the stationary WFs at each step, the total
expressions (\ref{eq.8}) transform into the corresponding
expressions for the problem of the particle propagating above the
barrier, coinciding with the solutions of the problems, studied in
\cite{McVoy.1967.RMPHA,Anderson.1989.AJPIA,Fermor.1966.AJPIA}.

We fulfill time analysis of the particle tunneling (or propagating)
relatively the barrier on the basis of the method of stationary
phase \cite{Maydanyuk.2000.UPJ,Olkhovsky.1992.PRPLC,
Olkhovsky.1997.JPGCE,Maydanyuk.1999.JPS}.
For obtaining of the tunneling and reflecting times it is
needed to know the expressions of the incident, transmitted and
reflected WPs relatively the barrier. In their calculation the
method of multiple internal reflections has shown itself as the
most effective in a comparison with other known methods.
This perspective is not visible in solution of the one-dimensional
problem of the particle tunneling through the rectangular barrier,
but is shown brightly in the solution of the problems of tunneling
through the barriers of the more complicated forms.
For the one-dimensional problem with the rectangular barrier
(\ref{eq.1}) one can find by the method of multiple internal
reflections both the components of the incident, transmitted and
reflected WPs (see~(\ref{eq.7}) and (\ref{eq.8})), calculate many
other $n$-multiple WPs, and determine on their basis many other
detailed time characteristics.

Now we formulate briefly an approach for a space-time description
of $\alpha$-decay of a nucleus, based on the method of multiple
internal reflections. We consider the decay process on the basis
of the quantum mechanical one-particle model of the decay of a
compound system with use of the barrier of the following form:
\begin{equation}
  V(r) = 
  \left\{
  \begin{array}{crcl}
    -V_{0}, & \mbox{ at } & r<R_{1}       & \mbox{ (region I)};  \\
    V_{1},  & \mbox{ at } & R_{1}<r<R_{2} & \mbox{ (region II)}; \\
    0,      & \mbox{ at } & r>R_{2}       & \mbox{ (region III)}.
  \end{array}
  \right.
\label{eq.10}
\end{equation}
Let's study a case, when the moment is $l=0$ and the decay occurs
at the energy levels, located below the barrier heights (i.~e. the
$\alpha$-particle tunnels through the barrier from the region I
into the region III). A general solution of the stationary WF has
the following form:
\begin{equation}
\begin{array}{ll}
  \varphi (r, \theta, \phi) =
    \displaystyle\frac{\chi(r)}{r} Y_{00} (\theta, \phi); &
  \chi(r) = 
    \left\{
    \begin{array}{lrc}
    A (e^{-ik_{1}r} + e^{ik_{1}r}),      & \mbox{ at } & r<R_{1}; \\
    \alpha e^{\xi r} + \beta e^{-\xi r}, & \mbox{ at } & R_{1}<r<R_{2}; \\
    S e^{ikr}, & \mbox{ at } & r>R_{2},
    \end{array}
    \right.
\end{array}
\label{eq.11}
\end{equation}
where $Y_{lm}(\theta, \phi)$ is the spherical function,
$k_{1} = \displaystyle\frac{1}{\hbar} \sqrt{2m(E+V_{0})}$,
$\xi = \displaystyle\frac{1}{\hbar} \sqrt{2m(V_{1}-E)}$.

According to the method, we describe the tunneling of the
$\alpha$-particle through the barrier with use of WP consecutively
by steps of its propagaition relatively each boundary, like the
one-dimensional problem.
At the first step, we study the beginning of the decay of the
compound nucleus. We consider the $\alpha$-particle, which
propagates inside the region I and is incident upon the internal
boundary of the barrier at point $r=R_{1}$. Here, we determine
the expression for the incident WP, which we use in definition of
the time moment of the $\alpha$-particle fall upon the barrier on
the inside (\emph{the time moment of the biginning of the decay of
the compound nucleus}, according to the method).
At the first step, the transmitted and reflected WPs start the
processes of the $\alpha$-particle leaving from the barrier region
and its delay inside the internal region I (the region of the
compound nucleus).

Further, we study the tunneling of the particle through the barrier
and its oscillationsÿinside the region I at the following steps.
Here, one can reduce any step of the WP propagation relatively the
barrier boundaries to one of 4 independent steps. With use of their
analysis the recurrence relations for the calculation of the
constant factors for the arbitrary step $n$ are constructed. The
total non-stationary WF in the each region, which takes into account
the multiple reflections (and transitions) of WP relatively the
barrier boundaries, can be present in the form of series of the
convergent and divergent WPs, which are coincided and can be
calculated by use of the recurrence relations obtained earlier.
In result, we find the incident, total transmitted and reflected
WPs relatively the barrier:
\begin{equation}
\begin{array}{lcl}
  \chi_{inc}(r,t) & = &
    \displaystyle\int\limits_{0}^{+\infty}
    g(E-\bar{E}) \Theta (V_{1}-E)
    A_{inc} e^{ikr - \frac{iEt}{\hbar}} dE, \\
  \chi_{tr}(r,t) & = &
    \displaystyle\int\limits_{0}^{+\infty}
    g(E-\bar{E}) \Theta (V_{1}-E)
    A_{inc} C_{tr} e^{ikr - \frac{iEt}{\hbar}} dE, \\
  \chi_{ref}(r,t) & = &
    \displaystyle\int\limits_{0}^{+\infty}
    g(E-\bar{E}) \Theta (V_{1}-E)
    A_{inc} C_{ref} e^{-ikr - \frac{iEt}{\hbar}} dE,
\end{array}
\label{eq.12}
\end{equation}
where
\begin{equation}
\begin{array}{lcl}
  C_{tr} & = &
    \displaystyle\frac{i4k_{1}\xi}
    {(k+i\xi)(k_{1}+i\xi + (k_{1}-i\xi) \exp{(2ik_{1}R_{1})})}
    \times \\
    & \times &
    \Biggl(1 -
    \displaystyle\frac{(i\xi-k)}{(i\xi+k)}
    \displaystyle\frac{i\xi-k_{1} - (i\xi+k_{1})\exp{(2ik_{1}R_{1})}}
                      {i\xi+k_{1} - (i\xi-k_{1})\exp{(2ik_{1}R_{1})}}
    \exp{(2\xi(R_{1}-R_{2}))}
    \Biggr)^{-1}; \\
\end{array}
\label{eq.13}
\end{equation}
\begin{equation}
\begin{array}{lcl}
  C_{ref} & = &
    \displaystyle\frac{C_{1}}{C_{2}}; \\
  C_{1} & = &
    \displaystyle\frac{i\xi-k_{1}}{i\xi+k_{1}}
    \exp{(2ikR_{1})}
    \Biggl(1 -
    \displaystyle\frac{(i\xi-k)(i\xi-k_{1})}{(i\xi+k)(i\xi+k_{1})}
    \exp{(2\xi(R_{1}-R_{2}))}
    \Biggr) + \\
    & + &
    \displaystyle\frac{4ik\xi(k-i\xi)}{(k+i\xi)^{2}(k_{1}+i\xi)}
    \exp{(2\xi(R_{1}-R_{2}) + 2ik_{1}R_{1})}; \\
  C_{2} & = &
    \Biggl(1 +
    \displaystyle\frac{k_{1}-i\xi}{k_{1}+i\xi} \exp{(2ikR_{1})} \Biggr)
    \Biggl(1 -
    \displaystyle\frac{(i\xi-k)(i\xi-k_{1})}{(i\xi+k)(i\xi+k_{1})}
    \exp{(2\xi(R_{1}-R_{2}))}
    \Biggr) + \\
    & + &
    \displaystyle\frac{4ik\xi(k-i\xi)}{(k+i\xi)^{2}(k_{1}+i\xi)}
    \exp{(2\xi(R_{1}-R_{2}) + 2ik_{1}R_{1})},
\end{array}
\label{eq.14}
\end{equation}
and $\Theta$-function satisfies to the condition:
\begin{equation}
  \Theta (\eta) = 
  \left\{
  \begin{array}{crc}
    0,     & \mbox{ at } & \eta<0, \\
    1,     & \mbox{ at } & \eta>0.
  \end{array}
  \right.
\label{eq.15}
\end{equation}
On the basis of these expressions one can calculate the
coefficients of penetrability and reflection of the
$\alpha$-particle relatively the barrier.

We fulfill the time analysis of the $\alpha$-decay of the compound
nucleus on the basis of papers
\cite{Maydanyuk.2000.UPJ,Olkhovsky.1992.PRPLC,Olkhovsky.1997.JPGCE,
Maydanyuk.1999.JPS}.
Here, we calculate the \emph{timeÿof the $\alpha$-decay of the
compound nucleus}, which we define as a difference between the
time moment of the $\alpha$-particle leaving from the barrier
region at point $r=R_{2}$ (\emph{the time moment of finishing of
the decay of the compound nucleus}, according to the method) and
the time moment of the beginning of the $\alpha$-decay of the
compound nucleus.
The time moment of the $\alpha$-particle leaving from the barrier
we define on the basis of the expression for the total transmitted
WP of the form (\ref{eq.12}) with taking into account of
(\ref{eq.13}). According to the calculations, the time of the
$\alpha$-decay of the compound nucleus has the form:
\begin{equation}
  \tau_{tr} =
    \hbar \displaystyle\frac{\partial}{\partial E}
    \arg C_{tr}.
\label{eq.16}
\end{equation}

In accordence with the analysis of the consequent transmissions
and reflections of the multiple WP relatively the barrier
boundaries, we note that the total WP leaves the region I not at
the time moment of the $\alpha$-decay of the compound nucleus, but
after the some period of time. One can suppose, that at the
beginning of the decay the particle makes oscillationsÿinside the
region I (this follows from the analysis of WF in this region).
A probability of its existence inside the region I decreases
evenly, but at the beginning of the decay process it is larger
than inside the regions II and III. We define the delay time as a
difference between the time moment of the falling of the total WP
inside the region I upon the internal boundary of the barrier at
point $r=R_{1}$ and the time moment of the beginning of the
$\alpha$-decay of the compound nucleus.

One can calculate the factor $A_{inc}$, used in (\ref{eq.12}), from
the normalization condition of WF, and it does not influence on the
values of the penetrability and reflection coefficients, and on the
values of the times of the $\alpha$-decay and delay.

In conclusion we note, that the approach for description of the
particle tunneling (and propagating) through the barrier on the
basis of the processes of the multiple internal reflections is
non-stationary and it reflects \emph{wave properties} of the
particle in its tunneling (or propagation). Applying of the method
of multiple internal reflections, which uses such approach, to the
description of the $\alpha$-decayÿof the spherical nuclei is the
further development of the method of multiple internal reflections,
which was applied earlier for solution of the problems of elastic
scattering of particles upon spherical nuclei and the one-dimensional
problems of tunneling of the particles through barriers of the
different forms. In the problem of the nuclear $\alpha$-decay with
its use the effect of the oscillations of the compound nucleus is
opened, which takes place at the beginning of the decay process.
We propose to take into accout such effect in calculations of
the decay time of the compound nucleus.

\bibliographystyle{h-physrev4}
\bibliography{MSar_eng}

%
%

\end{sloppypar}
\end{document}